# QREME – Quality Requirements Management model for supporting decision-making

Thomas Olsson[1] and Krzysztof Wnuk[2]

[1] RISE SICS AB, Lund, Sweden, `thomas.olsson@ri.se`
[2] Krzysztof Wnuk, DIPT, BTH, Karlskrona, Sweden, `krzysztof.wnuk@bth.se`

**Abstract.** **[Context and motivation]** Quality requirements (QRs) are inherently difficult to manage as they are often subjective, context-dependent and hard to fully grasp by various stakeholders. Furthermore, there are many sources that can provide input on important QRs and suitable levels. Responding timely to customer needs and realizing them in product portfolio and product scope decisions remain the main challenge.
**[Question/problem]** Data-driven methodologies based on product usage data analysis gain popularity and enable new (bottom-up, feedback-driven) ways of planning and evaluating QRs in product development. Can these be efficiently combined with established top-down, forward-driven management of QRs?
**[Principal idea / Results]** We propose a model for how to handle decisions about QRs at a strategic and operational level, encompassing product decisions as well as business intelligence and usage data. We inferred the model from an extensive empirical investigation of five years of decision making history at a large B2C company. We illustrate the model by assessing two industrial case studies from different domains.
**[Contribution]** We believe that utilizing the right approach in the right situation will be key for handling QRs, as both different groups of QRs and domains have their special characteristics.

## 1   Introduction

Quality Requirements (QRs, a.k.a. non-functional requirements, NFRs), defined as "attributes of or constraints on a system." [1], are ever-increasingly important [2, 3] but also challenging to handle. There are many challenges associated with QRs, e.g., insufficient product usability [4], project overruns, increased time-to-market [5], poor cost estimation or lower priority of quality compared to functionality [6] and poor validation of QRs [7, 8].

Extensive research was conducted in eliciting [2] representing and modeling QRs [7], leaving the areas of their realization and release planning greatly unexplored. At the same time, our previous work brings evidence that realizing QRs puts new demands on scoping and release planning [2, 9, 10], e.g., QRs often require more than one software release to be realized, top-down planning is not sufficient in many cases and there is a lack of support for executing product strategies based on QRs. Making decisions about what requirements to focus on is often called scoping. Scoping is usually performed by a product manager at a product level [11] and impacts portfolio strategy and product success [12]. Requirements scoping is a continuous activity that supports in translating the product strategies into a series of software releases [11].

Several researchers studied QRs and challenges associated with them. In our previous work, QRs appear to be unequally distributed within the same specification and the same company [10]. Ernst and Mylopoulos analyzed open source projects and concluded that there are large differences among projects and no clear correlation to the project age [13]. Concerning release planning, Ameller et al. report that most models provide "simple output for which requirements to implement in the next release" [14]. Others have identified an under-emphasis of product quality and difficulties in handling cross-cutting concerns across teams with agile methodologies [15].

In this work, we present QREME – Quality Requirements Management model for supporting decision-making about QRs. QRs are changing over time (even if you do not actively make decisions on them), QRs are always present in the software whether you have made explicit decisions on them or not and scoping is a continuous activity, scoping for QRs is different from traditional scoping. The main parts of QREME are: 1) the prominent roles and their responsibilities when making decisions on QRs, 2) the decision forums and how they are related, and 3) the strategic and operational levels for both product-related decisions and business intelligence related decisions. We focus on the following research question: ***How can we support portfolio and product decision makers with respects to QRs?*** The benefit of using QREME is a combination of more effective scoping decisions (making the right decisions) and a quicker response to changes in the marketplace and to software quality issues. It will also be easier to plan the improvement of QRs over several releases. As QREME is addressing the scoping of QRs, the specific way in which QRs are modeled and documented are as such impacting the use of QREME.

The paper is organized as follows: Section 2 introduces background and relevant related work. The research methodology is described in Section 3. The proposed model with relevant descriptions is found in Section 4 and two cases using the model is elaborated on in Section 5. Section 6 concludes the paper, including future work.

## 2    Background and Related Work

There are several definitions of QRs [1]. One implication of the definition we use in this paper of QRs as "attributes or constraints on a system" is that a QRs cannot exist without a corresponding functional requirement or (sub-) system. This, in turn, implies that a requirement or a system will always exhibit the attribute or constraint even if it is not explicitly specified. For example, a system always has a startup-time even if it is not explicitly expressed with a QR. In this paper, we use the term Quality Attribute (QA) as the abstraction of a specific QR., For example, start-up time is a QA and "the system should start in 2 seconds" a QR. Furthermore, we use the term Quality Level (QL) for the measurable level of a QR, in alignment with our previous work [2]. From the example, "2 seconds" is the QL.

The continuous nature of requirements scoping plays a vital role in bridging strategic product portfolio planning and associated release planning with operational scope decisions that need to be taken to adapt to unexpected changes [16]. However, linking business strategy to detailed planning is non-trivial [17]. The software product management literature [18] recognizes the strategic importance of QRs in setting the product strategy [19] but does not consider its particular nature during the product and release planning processes. Our previous empirical work shows that QRs should be incrementally delivered and the scoping process stretches over several product releases [9].

Traditional software development is typically done in a forward feeding and top-down manner, typified by the waterfall model [20]. In a forward feeding process, ideas or goals are the starting point and are broken down into requirements, later to be implemented and verified. In the end, the resulting product is evaluated against the original ideas and goals. Today, feedback-driven or bottom-up approaches are gaining momentum, often supported by data-driven approaches [21] or crowd-based approaches [22]. Objective data usage can remove subjectivity from the product managers [23]. In a feedback-inspired process, ideas and goals emerge from the actual usage, mostly through experimentation on alternatives to improve the product, and lastly evaluated against requirements and strategies whether the product is evolving in the right direction. However, for feedback driven approaches, is not clear which type of information is needed for scoping and how to achieve alignment among stakeholders. In our previous work, we saw a need to combine both forward and feedback processes [9]

Agile approaches such as Scrum [24] or the ideas with DevOps [25] end up somewhere in-between. This transition has substantial implications for software product strategies, product requirements engineering and product scoping. Increased flexibility in decision making that the above transformations bring puts more pressure on the synergy between strategic planning, product scoping, requirements management and realization. Incremental delivery of software gains importance and impacts release planning methods and processes [26].

Scoping decisions are often interdependent [27], continuously made [11] during different steps in the development process [27], in different forums [12], at several abstraction levels [28] and often not in a top-down fashion [9]. We studied release planning for QRs [2] while Carlshamre et al. focused on interdependencies among requirements in software release planning [29]. Berntsson-Svensson re-used the interdependency types suggested by Carlshamre et al. to study dependencies between QRs but without an apparent release planning angle [6]. Our work focuses on how to plan and deliver QRs across many releases, with each release taking the software closer to the fulfillment of the complete QR and desired QL.

# 3 Research methodology

We used Canonical Action Research (CAR) to develop the framework presented in this paper [30]. The focal point of CAR is a real-word problem that researchers attempt to address by combining scholarly observations with practical interventions using mostly interpretivist epistemology [30]. During one cycle, we continuously interacted with the environment under research and the subjects in this environment to reflect on the needs supported by the model.

**Problem investigation.** Previous work on analyzing decision patterns for quality requirements [9] have shaped the scope and goals of the current research. We have studied 4444 features from a period of 5 years from the beginning of a new product portfolio across many product and software releases. We combined decision history and document analysis with the interviews with key stakeholders involved in the decision-making process. Our main findings are: 1) QRs require planning across several releases, as they tend to require long lead-time and effort planning 2) some quality aspects (e.g. efficiency) were handled in a bottom-up fashion while other aspects (e.g. security) were driven from a top-down strategic process and 3) multiple strategies are required to have a responsive and aligned organization. The strong need for improved decision making about QRs was expressed during the interviews with the key stakeholders involved in the decision-making process.

**Treatment design.** During the development of QREME, we focused on creating QREME as "instrumental theory" that helps in generating coherent explanations and achieving understanding for decision making about QRs [30]. A clear need emerged early in the design process to handle both strategic and operational decision-making levels [27] as decisions on these levels are often interconnected. Moreover, the observations made from the in-depth analysis of 5 years of decision making about QRs confirmed that both feedback-loop and forward-loop are unsystematically used and needed to properly handle decision about QRs [9].

QREME was incrementally designed in a series of meetings where the authors discussed the versions and made changes and updates. Each new version of the model was critically evaluated and discussed in a workshop session among the researchers. Changes and updates were documented to enable traceability. The first version of QREME contained only the portfolio strategy and the product scope elements, based on empirical data [9] and related work [27]. After evaluations, it was decided that the core element of the feedback-loop is the product usage data that decision-makers need to continuously analyze and filter. Therefore, the analytics scope element was added to QREME. Next, the three decision forums were identified we named the input and output for each of the forums. Finally, in the last iteration, the roles involved in each decision forum were detailed.

**Treatment Implementation and Evaluation.** We evaluated QREME on two exploratory case studies from two companies developing software-intensive products but having different QR profiles. The evaluation consists of an assessment based on expert opinion on the companies' ways of working and which elements of QREME they are compliant with and would benefit from.

Company A focuses on user experience, performance and security as it develops software-intensive products of daily use for consumers. Company B, on the other hand, develops software-intensive products for B2B and is mainly concerned with performance, security, and maintainability. The products that company B develops have no user interface that the customers can interact with but collect digital images that can be analyzed in the software that combines it from several devices.

In the next phase, we will evaluate with companies and practitioners the underlying findings from [9] in other companies to ensure this is not unique the company used in that study. Furthermore, we will validate that QREME addresses the findings and is usable in a practical context.

### 3.1   Threats to validity

We discuss the validity threats according to the four perspectives on validity proposed by Yin [31] and some of the guidelines provided by Runeson and Höst [32].

*Construct validity* is concerned with establishing appropriate methods and measures for the studied phenomena or concepts. The empirical evidence that the framework is based on was collected from both the analysis of the decision-making logs and in interviews. This multiple-source evidence provides trustful catalog of observations that impacted the design decisions for the framework. Moreover, we worked inspired by CAR [30] where a theory is the focal point of generating coherent explanations of the studied phenomena and QREME can be considered as an instrumental theory of decision making about QRs.

*Internal validity* is concerned with uncontrolled confounding factors that may affect the studied causal relationships. The relationships between the selected decision strategies were anchored in the empirical data obtained in our previous study [9]. Still, a threat remains that when QREME is put into operation at other than studies industrial contexts, we may discover additional confounding factors that may affect the decision processes and therefore should be further incorporated into the framework.

*Reliability* is concerned with the degree of repeatability of the study. The framework creation process was continuously documented to enable traceability and analysis. The QREME creation process was inspired by CAR guidelines to ensure rigor during the iteration and collaboration between researchers and practitioners [30]. However, reliability of the interpretations made during QREME development could be questioned as this step of the process remains highly subjective. We took precautions to minimize subjectivity by discussing our interpretations with industry practitioners and between the authors and seeking most reliable explanations.

*External validity* remains the main concern of this work. QREME is based on an in-depth analysis of five years of decision making about QRs at a large company. Still, we cannot claim that this is a representative case of how all software-intensive product development companies deal with QRs. Therefore, the suitability of QREME must be further validated outside the two contexts described in the case studies to bring supporting evidence that the foundations of the theoretical framework remain strong for other contexts, product types, and requirements engineering processes.

## 4 Quality Requirements Management model (QREME) Supporting decision-making for quality requirements

The goal of QREME is to provide decision support for managing quality requirements, incorporating two highly interconnected processes: a top-down forward driven, and a bottom-up feedback process. QREME can be applied as an assessment instrument as well as to plan improvement activities for scoping of quality requirements.

### 4.1 The anatomy of QREME

QREME has two abstraction levels for decisions: a strategic level and an operational level [9, 27], see **Fig. 1**. At the strategic level, strategic product decisions are handled, such as deciding which quality aspect (QA) to address and what customer segments to focus on. At the operational level, decisions for individual products are handled, such as quality level (QL) for a specific QR for a specific release or analysis of usage data in a specific context. For example, a QR can be start-up time from powering on a device and the QL can be 10 seconds. The operational decisions are usually short-term and consider individual products and releases.

Furthermore, QREME separates scoping decisions on the products from decisions on data analysis. Product decisions are about what the products should realize and what data to utilize in the experiments.

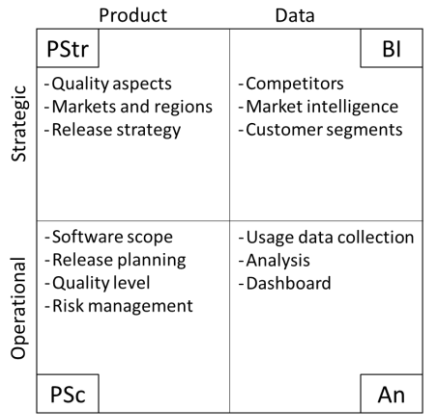 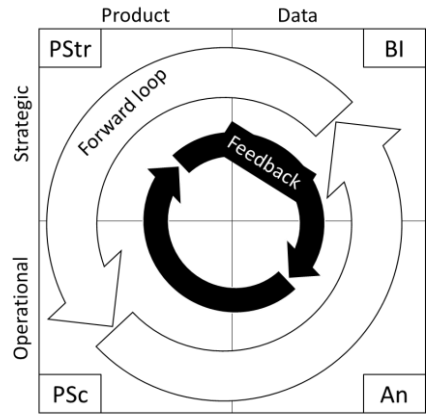

**Fig. 1.** The conceptual overview of QREME, with PStr = Portfolio Strategy, PSc = Product Scope, BI = Business Intelligence and An = Analytics. **Fig 1a** summarizes types of decisions in the different areas of QREME and **Fig 1b** illustrates the two loops.

This results in four scope decision areas: Product portfolio strategy (PStr), Product Scope (PSc), Business intelligence (BI) and Analytics (An), see **Fig. 1a**. QREME also distinguishes between (product-)planning-driven decisions (forward-loop) and data-driven decisions (feedback-loop), see **Fig. 1b**. For the two highly interconnected loops, the feedback-loop is usually faster than the forward-loop. Both loops traverse the four scope decision areas in opposite directions and at different speeds.

### 4.2 Scope decision areas

To achieve both a structured process in refining a long-term roadmap as well as an agile and short response-time to changes in the market, all four decision areas need to have a certain level of autonomy, independent input and possibility to influence each other.

**The PStr area** concerns strategic product decisions such as quality aspects, markets and release strategy. The decisions are on a strategic level [27] and should embody a company's strategy for the product(s) or portfolio. A **portfolio manager** is typically the main decision maker [33]. Decisions will outline portfolio-wide direction regarding which QAs to focus on and how individual products show relate to this. PStr decisions should be reviewed on a quarterly or half-year interval. The main decision forum is **the product portfolio strategy forum**. The decisions are typically summarized in informal natural language as a presentation file or a short document. The portfolio manager mainly interacts with the product manager and the business intelligence manager for scope decisions, cf. **Fig. 2**. Besides the roles directly involved in the decisions, the portfolio manager takes input from executive management, marketing manager, key account managers, etc.

**The BI area** is also on the strategic level. BI decisions concern which competitors to monitor, which market data to collect and how to divide the customers into the relevant customer groups, etc. It can also be areas where the company wants to experiment (e.g., through A/B testing) rather than performing a (traditional) upfront requirements analysis. A **business intelligence manager** is the main decision maker for BI. Decisions should outline relevant BI data to ensure adequate coverage. Similar to PStr decisions, BI decisions should be reviewed and updated on a quarterly or half-year interval. The main decision forum is **the business intelligence decision forum**. The BI data is presented with graphs and numbers but in informal documents or presentations. Besides interacting with PStr and An regarding scope decisions, input comes from marketing manager, competitive intelligence, sales, etc.

The BI manager interacts with the portfolio manager on the strategic level and the analytics manager on the operational level (cf. **Fig. 2**). The BI manager also interacts much with, e.g., marketing managers, customer services and external companies to collect competitive intelligence.

**The PSc area** operational decisions (see **Fig. 1a**) target QL for a specific QR and the realization strategy in the coming releases. A **Product Manager** is responsible for PSc decisions [33]. Depending on the development context and release interval, PSc decisions could be made a weekly or monthly interval, or continuously. The main decision forum is the **product scope decision forum**. In an agile context, a more informal continuous dialogue in the team replaces the formal product scope decision forum. PSc decisions are on an operational level and in a semi-structured format e.g. in an issue handling tool, decision database or spreadsheet backlog. The product manager receives the portfolio strategy from the portfolio manager and product usage data from the An-alytics manager (see **Fig. 2**). The product manager also interacts with key account managers, internal stakeholders, such as subject area experts and the development organization, and external stakeholders, such as customers and key account managers.

The **An area** decisions concern the product usage data collection and analysis. If a company is utilizing experimentation or beta-testing, decisions on how many experiments to run and how closely to monitor the product usage is an An decision. Especially important is to be wary of the amount of data generated, as collecting usage data can result in the copious amount of data. An **Analytics manager** oversees the **An area** decisions. Analytics consists of one part focused on instrumentation and the actual usage data collection and one part of the analysis and presentation of the data. Decisions on which usage data to collect are made daily or weekly. Decisions are made either in centralized **Analytics decision forums** or distributed in different development teams. Analytics decisions are presented alongside with the rich and highly structured data. The analytics manager receives the product scope from the product manager as well as a usage data scope from the BI manager (see **Fig. 2**). The analytics team also interacts closely with the development team for the instrumentation and actual data collection. Competitor devices can also be used to compare specific measurements with.

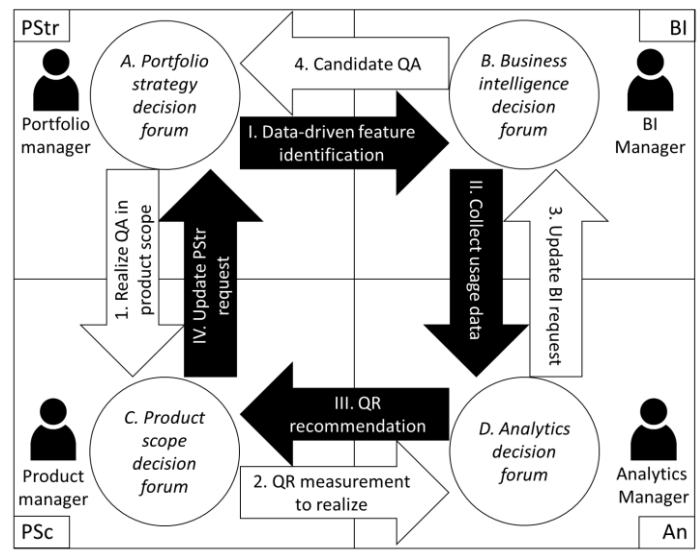

**Fig. 2.** Interactions among the four scope decision areas in the forward- and feedback-loops.

### 4.3 The interaction between roles and decision forums in QREME

The four scope decision areas are connected and impact each other through the two loops, as outlined in the previous section. **Fig. 2** outlines how decisions from different decision forums are connected to each other.

**Table 1** Alternatives for scope decisions for the different forums

| Forum | Forward-loop | Feedback-loop |
|---|---|---|
| A | **1. Realize QA in product scope** - The PStr can decide to have the PSc refine the decisions. This is a typical refinement of a QA to QRs, suitable when the market needs are well understood or when there is no comparable experience to learn from (e.g. radical innovation). | **I. Data-driven feature identification** – The PStr can decide to have BI analyze the actual needs. Instead of up-front QAs refinement, experiments determine the appropriate QLs. This is suitable when it is difficult to upfront estimate QLs or there is an opportunity to incrementally improve a QA in a data-driven manner without any change to QAs as such. |
| B | **4. Candidate QA** – The decisions which QAs to improve as identified by BI can be send to PStr as a candidate QAs included in PStr. It can be market trends or QLs in the existing software which stands out in BI and it not represented in the PStr. The forward-loop is suitable when the product QLs are known and there is an identified gap in the PStr. | **II. Collect usage data** – Decisions on which is the relevant QAs to collect refined data can be request of An. Based on market and competitor analysis and input from PStr, the BI identifies QAs which need clarification on regarding QLs in the software. This is appropriate when the QLs in the software are unknown and the QAs are part of the PStr. |
| C | **2. QR measurement to realize** – PSc decides which QRs to implement. In the forward-loop, PSc requests An to collect usage data for the relevant QRs being implemented. This is a kind of refinement in terms of collecting data for defined QLs. The forward-loop from PSc to An for scoped decisions is appropriate when the QRs and their QL is in line with the portfolio strategy. | **IV. Update PStr request** – If the product manager finds gaps in what they want to highlight to PSc, then the product manager can decide to send a QR to request PStr to update PStr with respect to the QR. This can happen when there is feedback from An on gaps or input from other stakeholders which PSc would like to include in the scope. The feedback-loop usage is appropriate when there is a discrepancy from the needs of the PSc and the PStr. |
| D | **3. Update BI request** – Based on usage data analysis, An can request to update BI. This can be if QLs and QRs are identified as relevant for An but this is not in line with the current BI scope. Utilizing the forward-loop from An to BI is appropriate when the BI need to include product usage data currently not covered by the BI strategy, as a result of the forward-loop from PSc. | **III. QR recommendation** –The An decision forum provides recommendations to the product manager on suitable QLs for different QRs as well as if there are specific QRs which need attention. The QAs might be identified in BI in the feedback-loop or be a feedback on QLs coming from PSc. The analytics manager should use the feedback-loop QAs in the strategy or when the customers express their strong dissatisfaction about QRs. |

The forward-loop (counter-clockwise in the figure) is characterized by top-down flow where PStr decisions and extracted into PSc decisions that are realized in software [19]. Customer sentiment and sales data are reported back to the portfolio management. The forward-loop is often exercised by bespoke or MDRE (Market-Driven Requirements Engineering) companies where the development is either performed in-house or regulated by a contract. In these situations, it is possible to work with our framework to create systematic information exchange among the different decision processes.

The feedback-loop (clock-wise) is constructed based on the assumption that a software-intensive company has access to product usage data [21]. As a result, instead of having upfront investments to analyze and synthesize a scope, inspiration is taken from the product usage or other sources (e.g., social media), though both understanding as well as exploring changes. Based on product usage data analysis, improvements are identified and made part of the product scope for implementation. The resulting product scope is evaluated in the portfolio strategy.

We assume that no organization uses only forward- or feedback-loop. Rather, they tend to favor one of the loops without sufficient synergy between them. For example, information flow between the An and PSc need to be efficient. Low efficiency of this information flow may result in long lead-times, e.g., when the product usage data is not promptly integrated into the PStr forum via either PSc or BI. Moreover, if a QA is not considered to be relevant in PStr, the information night never reaches PSc.

There are four interactions in the forward-loop (labelled 1-4) and four interactions in the feedback-loop (labelled I-IV) among the decision forums (labelled A-D), see **Fig. 2**. We make two assumptions:
1. Decisions (and development) are made in all the forums continuously.
2. Decisions are made individually.

Hence, we are not considering the situation, e.g. where a requirements specification is prepared and finalized and then sent onwards in the process. Furthermore, there is no explicit beginning or end in the loops as most software-intensive companies work with existing portfolios and products and seldom create new portfolios. **Table 1** outlines a guideline to choose whether to use the forward- or feedback-loop.

### 4.4 Tailoring QREME

The ideal model of QREME as described in the previous sections need to be adapted to the specific organization and their needs. In tailoring QREME, the central aspect to consider is to cater for the two loops and leverage from the different characteristics; the forward-loop with long-term planning and the feedback-loop with shorter lead-time to changes in the market and the software. However, the specific roles or decision forums are not crucial to have as in the ideal model. Instead, the critical aspect is to be aware of the different types of decisions and map the roles to the ones in the organization.

## 5 Two exploratory case studies

In this section, we present two case studies that provide experiences from applying QREME. The application consists of using QREME in an expert assessment on the companies' current decision processes related to scoping of QRs

### 5.1 Case A: Consumer device products for a global market

**Company A** develops software-intensive products for B2C for a global market. Development is performed in a cooperative manner with other companies, sometimes called Software Ecosystems (SECO) [34]. Substantial investment is made in the software developed for the dedicated hardware. QAs play a crucial role in product success as well as customer purchase decisions. We performed an extensive longitudinal study of the decision patterns [9], which lay the underlying rationale for QREME.

**Case:** One of the observations was an issue with battery performance. The company releases several products per year. The software is updated several times over the lifecycle and up to 2 years after the product first reaches the market. Even though software today is a significant part of the engineering efforts, the underlying hardware platform brings substantial opportunities and limitations regarding the possible quality aspects. This "hardware legacy" is still visible regarding processes and culture, leading to a prevalence of the forward-loop.

**Portfolio strategy.** In the portfolio strategy decision forum, it was decided an overall target for battery performance and it had been the same for several product generations. However, despite reports of not meeting the target level, there were no actions on a portfolio level. Given the legacy of hardware development, there is a strong focus on the hardware side of the portfolio for the new products to reach the market and less focus on the software updates for existing products. The software product managers are not represented (cf. **Fig. 2**). Furthermore, the BI manager role is not present. Instead, the software organization undertakes ad-hoc measurements of the product usage. Albeit product managers repeatedly highlighting battery problems, the portfolio strategy decision forum failed to timely and appropriately react.

When assessing this case using the framework, the feedback-loop, mainly with IV. in **Fig. 2**, is the most prominent problem in the portfolio strategy decision forum causing a delayed updated of the portfolio strategy. Furthermore, once the portfolio strategy was updated, there was still a focus on the forward-loop. We believe that it might have been more effective to employ a feedback-loop, see interaction I in **Fig. 2**, as the setting of an unrealistic QLs in the portfolio strategy, had previously shown to be ineffective. Hence, instead of using a feedback-loop, the battery performance should be improved until there is a positive sentiment rather than fulfilling a somewhat random number.

**Product scope.** The product manager got input from the portfolio manager to achieve a specific target QL for battery performance. However, it was one of many aspects needed fulfillment and was when the problems started to occur not prioritized among other features and QRs. There was also a strong focus on the products' introduction to the market and less focus on the software updates, limiting the ability to work with the product scope for later releases. This is further complicated by the fact that both the users' behavior (what they are doing) as well as the execution environment (the network) influence QAs. Hence, setting appropriate QLs upfront is challenging.

The feedback-loop from development to product scope work well regarding the framework (cf. III. in **Fig. 2**). In the product scope decision forum, the product manager and representatives from the development organization are present. This creates a strong relationship and quite well working forward-loop and relative well feedback-loop. However, the product manager had difficulty to act on feedback, as the portfolio manager expected the portfolio strategy to be prioritized and as so often it caused an over-scoping. Furthermore, there was no explicit data scope role and no strong tradition to experiment. Because the forward-loop preference from the portfolio and focus on the first release of the products to the market, this also caused an over-scoping for the first release and down-prioritizing of software updates.

**Analytics scope.** There was no explicit analytics manager role as intended in our framework. Instead, the development organization, through the project manager performed some of the usage data collection tasks. However, there was no tradition or explicit ambition to experiment or test improvements on parts of the consumer base and form the analytics decision forum either.

**BI scope**. The BI is much focused on external input such as market and competitor data and less on internal data such as usage and customer services data. There is a strong focus on pre-release of new products and their perception as they are first introduced to the market. There is much less focus on monitoring the perception of the (software) products during the whole lifecycle.

There is a gap in the feedback-loop in that the communication in I and II (cf. **Fig. 2**) are mostly missing. Even if the portfolio strategy is used for BI in general, it is not used to understand specific QAs in the products. Furthermore, there is little or no direction from BI to the teams collecting usage data and performing analysis thereof. This case a fragmented picture and lack of actionable intelligence in a strategic level.

To summarize, there is a strong focus on the forward-loop, i.e., 1-4 in **Fig. 2**. The main information presented to the portfolio manager is related to general performance of the products including specific QRs. Specific suggestions were communicated from the analytics team to the product manager (III. in **Fig. 2**). However, the feedback-loop from the portfolio manager to BI was effectively non-existent. Hence, there was no ambition to experiment and measure on software and incrementally update it in a data-driven way. Instead, it was expected that analytics is driven in a forward-loop manner.

The main benefit of improving the feedback-loop is expected to be a significantly shorter lead-time to adapt to customer expectations and changes in the market. Furthermore, sometimes decisions are made early in the process without real data. By introducing a clearer feedback-loop and daring to leave details to a later stage, more appropriate QLs will be implemented (neither too conservative nor over-shooting the target) which will in the end mean more effective use of development resources.

### 5.2   Case B: B2B product developing company

**Company B** develops software-intensive products for B2B contexts for a global market in a market-driven manner. In this case, we analyze performance requirements just as in Case A. This illustrates a different approach to handling the QRs and how QREME can support it.

**Case**: Company B is one of the world's leaders in its market segment despite having no official requirements database and only lightweight and informal requirements management processes. The requirements are often expressed in a comparative way as "benchmarking", e.g., "Product x should be as Product y, but better" and "The new version of the software must not be worse than the last version". This way of expressing requirements combined with test-driven development methods created a very strong feedback-loop based on continuous validation of the product behaviors by engineers.

**Portfolio strategy.** The portfolio strategy forum decision mainly focuses on new functionality and associated technical novelties. QRs and expected QLs are well understood and acknowledged but rarely quantified or explicitly documented. The leading requirements specification technique is to express the requirements about current or previous software capabilities. This creates issues in translating the strategy into objective QRs and the product scope.

Despite the best efforts, the forward-loop (1 in **Fig. 2**) is not sufficiently established to perform refinement into features with sufficient QRs and QLs and to later assess strategy fulfillment. On the other hand, the data-driven feature identification (II) works well when customers signal insufficient QLs that are escalated into the portfolio strategy decision forum. The role of the portfolio manager is not present in the organization as the responsibility falls between the executive management and product managers who have a limited responsibility for their products.

**Product scope.** The refine of functional features in the product scope decision forum worked well but not for QRs. Due to lack of strategic guidelines and "benchmarking as requirements", the product manager could not effectively communicate with experts and developers. The "benchmarking as requirements" had to be combined with feature/product usage data. However, in this case, usage data was replaced by test data obtained from the lab. Software developers or testers ran the previous products on example use cases and measured the current QLs for performance and other quality aspects. No additional product testing and product usage data was generated leaving little guideline or support for scoping decisions. Regarding QREME, the feedback from analytics team to product manager (III in **Fig. 2**) works well. However, the forward communication (2) is mostly lacking, which makes analytics mostly reactive.

**Data scope.** Developers, testers and often requirements engineers perform product tests to obtain reliable QLs. The product usage data arriving from the customers is only analyzed from the functional requirements viewpoint. The analytics manager role is not clearly established and clear data usage input challenges are not maintained. Upon incoming feature requests, this forum can only answer by providing QLs of previous products that can form a baseline for improvement suggestions. Potential feature recommendations are mostly functionality centered and lack clear QLs.

**BI scope.** The BI is much focused on external input from the market and competitors and direct customer data channels are not available for the company. The company sells its products via retailers who take the responsibility for hardware and software installations. Moreover, data is often secured by the customers and special permissions or legal documents are required to obtain it, e.g. by authorities. The company runs various products and software versions in the lab to obtain product usage data and to measure performance levels for products sold to the customers.

QREME highlights a need for an explicit role for the analytics manager. Furthermore, since there is no culture of experimenting or collecting product usage data, there is a need for education and training. QREME also emphasizes the need for more explicit channels and roles for portfolio management, to be able to quicker make changes relevant for the customers and markets.

The main benefit from applying QREME in this case is improving the forward-loop and increasing the synchronization effect between the feedback- and the forward-loop. Establishing the forward-loop and associated roles should mitigate the issues in translating the strategy into QRs and the product scope. Moreover, this should enable more proactive QRs definition rather than reactive response to customer dissatisfaction.

## 6 Conclusion and Future work

In this paper, we addressed the research question for how to support decision-making for QRs. Based on related work and our empirical work on understanding the decision patterns for QRs, we propose a decision-making model to align roles and forums for QR decisions combining a forward-loop and feedback-loop on strategic and operational levels. The focal points of the QREME model introduced in this paper are the two loops and the group decision making forums.

We applied QREME into exploratory case studies where we performed two assessments of how two companies make decisions for QRs. Using QREME, we identify several challenges in handling QRs that the companies should focus on addressing; namely an over-emphasis on a forward-loop and lack of common direction for QRs. We see a potential to shorten lead-times to react to changes in the market and customer expectations as well as a more efficient use of development resources with more accurate setting of QLs and therefore not wasting resources.

QREME has not yet been rolled out for daily operational work at any of the studied companies. Therefore, we plan to integrate QREME into the daily requirements operations and decision making at the partner companies and measure the long-term impact of it. Besides that, we also see a need to understand in more detail the contextual factors influencing the choice of the forward-loop and the feedback-loop, especially for innovation and the strategic portfolio decisions but also product lifecycle and market maturity. Finally, we plan to integrate various requirements abstraction levels of requirements into the model and detailed requirement levels for the decision forums.

We believe that the improved understanding of QRs, specifically regarding the feedback-loop, can have a positive influence on getting companies to emphasize on QRs. In our experience, the development organization is often aware of the QRs, but at the same time, portfolio and product management typically do not drive improvement of QRs. By introducing a clearer feedback-loop and thus making the QRs explicit, both the understanding that addressing the QRs in the software takes up development resources and user experience of the product is improved. This, we speculate, can help to create a foundation for an overall clearer prioritization of QRs at all levels and both in the forward-loop as well as the feedback-loop.

**Acknowledgements.** We want to thank all the participants in the interviews. This work is supported by the IKNOWDM project (20150033) from the Knowledge Foundation in Sweden.

## References

1. Glinz, M.: On non-functional requirements. In: IEEE International Conference on Requirements Engineering. pp. 21–26, Piscataway, NJ, USA (2007).
2. Regnell, B., Berntsson-Svensson, R., Olsson, T.: Supporting roadmapping of quality requirements. IEEE Softw. 25, pp. 42–47 (2008).
3. Regnell, B., Berntsson-Svensson, R., Wnuk, K.: Can we beat the complexity of very large-scale requirements engineering? In: Requirements Engineering: Foundation for Software Quality. pp. 123–128, Montpellier, France (2008).
4. Ebert, C.: Putting requirement management into praxis: dealing with nonfunctional requirements. Inf. Softw. Technol. 40, pp. 175–185 (1998).
5. Cysneiros, L.M., Leite, J.C.S.D.P.: Nonfunctional requirements: from elicitation to conceptual models. IEEE Trans. Softw. Eng. pp. 30, 328–350 (2004).
6. Svensson, R.B., Gorschek, T., Regnell, B., Torkar, R., Shahrokni, A., Feldt, R.: Quality Requirements in Industrial Practice - An Extended Interview Study at Eleven Companies. IEEE Trans. Softw. Eng. 38, 923–935 (2012).
7. Chung, L., Nixon, B.A., Yu, E., Mylopoulos, J.: Non-Functional Requirements in Software Engineering. Springer USA (2000).
8. Mylopoulos, J., Chung, L., Nixon, B.: Representing and Using Nonfunctional Requirements: A Process-Oriented Approach. IEEE Trans. Softw. Eng. 18, pp. 483–497 (1992).
9. Olsson, T., Wnuk, K., Gorschek, T.: Decision patterns for Quality Requirements: An Empirical Study, submitted to Journal of Systems and Software.
10. Berntsson-Svensson, R., Olsson, T., Regnell, B.: An investigation of how quality requirements are specified in industrial practice. Inf. Softw. Technol. 55, pp. 1224–1236 (2013).
11. Wnuk, K., Kollu, R.K.: A Systematic Mapping Study on Requirements Scoping. Proc. 20th Int. Conf. Eval. Assess. Softw. Eng., Limerick, Ireland (2016).
12. Regnell, B., Brinkkemper, S.: Market-driven requirements engineering for software products. Eng. Manag. Softw. Requir. Berlin Heidelb. Pp. 287–308 (2005).
13. Ernst, N.A., Mylopoulos, J.: On the perception of software quality requirements during the project lifecycle. In: Requirements Engineering: Foundation for Software Quality. pp. 143-257, Essen, Germany (2010).
14. Ameller, D., Farré, C., Franch, X., Rufian, G.: A Survey on Software Release Planning Models. Proc. 20th Int. Conf. Eval. Assess. Softw. Eng., Limerick, Ireland, pp. 48-65 (2016).
15. Mohagheghi, P., Aparicio, M.E.: An industry experience report on managing product

quality requirements in a large organization. Inf. Softw. Technol. 88, pp. 96-109 (2017).
16. Wnuk, K., Gorschek, T., Callele, D., Karlsson, E.A., Åhlin, E., Regnell, B.: Supporting Scope Tracking and Visualization for Very Large-Scale Requirements Engineering-Utilizing FSC+, Decision Patterns, and Atomic Decision Visualizations. IEEE Trans. Softw. Eng. 42, pp. 47–74 (2016).
17. Komssi, M., Kauppinen, M., Töhönen, H., Lehtola, L., Davis, A.M.: Roadmapping problems in practice: value creation from the perspective of the customers. Requir. Eng. 20, pp. 45–69 (2015).
18. Kittlaus, H.B., Clough, P.N.: Software Product Management and Pricing. Springer Berlin (2009).
19. ISPMA: Software Product Management - Foundation Level v.1.2. 1–39 (2014).
20. Royce, W.: Managing the Development of Large Software Systems. Proceedings, IEEE WESCON. pp. 1–9 (1970).
21. Bosch, J.: Speed, Data, and Ecosystems: The Future of Software Engineering. IEEE Softw. 33, pp. 82–88 (2016).
22. Groen, E.C., Doerr, J., Adam, S.: Towards Crowd-Based Requirements Engineering A Research Preview. In: Requirements Engineering: Foundation for Software Quality, pp. 247–253, Essen, Germany (2015).
23. Johansson, E., Bergdahl, D., Bosch, J., Holmström Olsson, H.: Requirement Prioritization with Quantitative Data - A Case Study. In: International Conference on Product-Focused Software Process Improvement. pp. 89–104. Bolzano, Italy (2015).
24. Schwaber, K.: Agile project management with Scrum. Microsoft Press (2004).
25. Hüttermann, M.: DevOps for Developers. Springer (2012).
26. Ruhe, G., Saliu, M.O.: The art and science of software release planning. IEEE Softw. 22, pp. 47–53 (2005).
27. Aurum, A., Wohlin, C.: The fundamental nature of requirements engineering activities as a decision-making process. In: Inf. and Software Technology. pp. 945–954 (2003).
28. Gorschek, T., Wohlin, C.: Requirements abstraction model. Requir. Eng. 11, pp. 79–101 (2006).
29. Carlshamre, P., Sandahl, K., Lindvall, M., Regnell, B., Natt och Dag, J.: An Industrial Survey of Requirements Interdependencies in Software Product Release Planning. Proc. Fifth IEEE Int. Symp. Requir. Eng. pp. 84–92 (2001).
30. Davison RM, Martinsons MG, Ou CXJ.The roles of theory in canonical action research. MIS Quarterly. 36, pp. 763-786 (2012).
31. Yin, R.K.: Case Study Research . Design and Methods. Sage Publications (2003).
32. Runeson, P., Höst, M.: Guidelines for conducting and reporting case study research in software engineering. Empir. Softw. Eng. 14, pp. 131–164 (2009).
33. Kittlaus, H.-B., Fricker, S.A.: Software product management : the ISPMA-compliant study guide and handbook. Springer Berlin (2017).
34. Jansen, S., Cusumano, M.A., Brinkkemper, S.: Software ecosystems : analyzing and managing business networks in the software industry. Edward Elgar Publishing (2013).